\documentclass[runningheads]{llncs}

\usepackage[T1]{fontenc}
\usepackage[utf8]{inputenc}

\usepackage{amsmath, amssymb}
\usepackage[english]{babel}
\usepackage{graphicx}
\usepackage{xspace, url, verbatim, array, float}
\usepackage[mathscr]{eucal}
\usepackage{cite}
\usepackage[version=3]{mhchem}
\usepackage{algpseudocode}
\usepackage{multirow}

\usepackage{dcolumn}
\newcolumntype{M}{D{.}{.}{6}} 
\newcolumntype{d}[1]{D{.}{.}{#1}}
\newcolumntype{C}{>{$}c<{$}}
\newcolumntype{R}{>{$}r<{$}}

\newcommand\Nset{\mathbb{N}}
\newcommand\Rset{\mathbb{R}}
\newcommand\WITH{\,:\,}
\DeclareMathOperator{\modulo}{\,mod\,}

\newcommand\etal{\emph{et~al.\@}\xspace}
\newcommand\naive{na\"ive\xspace}
\newcommand\decomposable{\mathit{decomposable}}
\def\phi{\varphi}
\newcommand\mumax{\mu_{\text{max}}}

\begin{document}

\title{Faster mass decomposition}

\author{Kai D\"uhrkop \and Marcus Ludwig \and Marvin Meusel \and Sebastian
  B\"ocker}

\institute{Chair for Bioinformatics, Friedrich Schiller University, Jena,
  Germany, \url{sebastian.boecker@uni-jena.de}}

\date{\today}

\maketitle


\begin{abstract}
Metabolomics complements investigation of the genome, transcriptome, and
proteome of an organism.  Today, the vast majority of metabolites remain
unknown, in particular for non-model organisms.  Mass spectrometry is one of
the predominant techniques for analyzing small molecules such as metabolites.
A fundamental step for identifying a small molecule is to determine its
molecular formula.

Here, we present and evaluate three algorithm engineering techniques that
speed up the molecular formula determination.  For that, we modify an
existing algorithm for decomposing the monoisotopic mass of a molecule.
These techniques lead to a four-fold reduction of running times, and reduce
memory consumption by up to $94\,\%$.  In comparison to the classical search
tree algorithm, our algorithm reaches a 1000-fold speedup.
\end{abstract}


\section{Introduction}

Metabolomics complements investigation of the genome, transcriptome, and
proteome of an organism~\cite{last07towards}.  Today, the vast majority of
metabolites remain unknown, and this is particularly the case for non-model
organisms and secondary metabolites: for many organisms, there is a striking
discrepancy between the number of identified metabolites and the prediction
of secondary metabolite-related biosynthetic pathways through recent
genome-sequencing results, see for example~\cite{cortina12myxoprincomide}.
The structural diversity of metabolites is extraordinarily large, and much
larger than for biopolymers such as proteins.  In almost all cases, we cannot
deduce the structure of metabolites from genome sequences, as it is done with
proteins.  Mass spectrometry (MS) is one of the two predominant experimental
analysis techniques for detecting and identifying metabolites and other small
molecules, the other being nuclear magnetic resonance (NMR).  The most
important advantage of MS over NMR is that it is orders of magnitude more
sensitive, making it the method of choice for medium- to high-throughput
screening applications \cite{last07towards}.  Newly identified metabolites
often serve as leads in drug design~\cite{li09drug}, in particular for novel
antibiotics~\cite{cooper11fix}.

In a mass spectrometry experiment, we measure the mass-to-charge ratios
($m/z$) of a peak, corresponding to an ion of the intact molecule or its
fragments.  Here, we omit the analysis of the charge state and assume that
the mass $m$ of the ion is known: In most cases, small molecules receive only
a single charge, and other cases can be either detected by analyzing the
isotope pattern on the metabolite, or simply by iterating over the few
possible charge states.

One of the most basic --- but nevertheless highly important --- steps when
analyzing a molecule, is to determine its molecular formula.  We note in
passing that mass spectrometry does not record the mass of the uncharged
molecule but rather the mass of the corresponding ion; for the sake of
clarity, we ignore this difference in the following.  Common approaches
first compute all candidate molecules with mass sufficiently close to a peak
mass in the measured spectrum, using an alphabet of potential elements. In a
second step, additional information is used to score the different candidate
molecules, for example using isotope pattern or fragmentation pattern
information.  The identified molecular formulas may then serve as a basis for
subsequent identification steps.  The problem of decomposing peak masses lies
at the core of practically every approach for the interpretation of small
molecule MS data that does not directly depend on a spectra library: See for
example \cite{boecker08towards, boecker09sirius, rasche11computing,
  stravs13automatic, meringer11msms, rojas-cherto11elemental,
  jarussophon09automated, pluskal12highly}.

First approaches for decomposing masses date back to at least the
1970's~\cite{robertson77massform, dromey80calculation}, where the \naive
search tree algorithm described below is mentioned for the first time.
Running times of this algorithm are often prohibitive, particularly for large
alphabets of elements.  F\"urst \etal~\cite{fuerst89computer} proposed a
faster decomposition algorithm which, unfortunately, is limited to the four
elements \ce{CHNO}.  For integer-valued masses, the problem is closely
related to unbounded integer knapsacks~\cite{martello90knapsack}.  Here, an
algorithm that works for arbitrary alphabets of elements is ``folklore'' in
computer science, and can solve the problem in pseudo-polynomial running
time~\cite{martello90knapsack}.  B\"ocker and Lipt\'ak
\cite{boecker05efficient, boecker07fast} presented an algorithm that
requires only little memory and is swift in practice.  Decomposing
real-valued masses using the integer-mass approaches was introduced
in~\cite{boecker09sirius}.  See also the
review~\cite{scheubert13computational}.

In this paper, we present three algorithm engineering techniques to speed up
the decomposition of peak masses in practice.  First, we replace the
recursive decomposition algorithm by an iterative version that mimics the
recursive enumeration, but is faster in practice.  Second, we show how to
minimize rounding error accumulation when transforming real-valued masses to
their integer-valued counterparts.  Finally, we modify the algorithm
from~\cite{boecker07fast} to decompose intervals instead of single masses,
based on ideas from~\cite{agarwal12stoichiometry}.  Together, these
improvements result in $4.2$-fold decreased running times, compared to the
previously fastest approach~\cite{boecker09sirius}.  We evaluate this on four
experimental datasets.


\section{Preliminaries} \label{sec:prelim}

In the following, let $a'_1, \dots, a'_k \in \Rset_{>0}$ denote the masses of
our alphabet~$\Sigma$, see Table~\ref{tab:masses} for masses of
elements.\footnote{For readability, we will denote the real-valued masses by
  $a'_j, l', u'$ and the integer-valued masses by $a_j, l, u$.}  We usually
assume that these masses are ordered and, in particular, that $a'_1$ is
minimum.  We want to decompose the mass of a peak in a measured spectrum, but
we have to take into account measurement inaccuracies.  To this end, we
assume that we are given an interval $[l',u'] \subseteq \Rset$, and want to
find all \emph{decompositions} $c = (c_1, \dots, c_k) \in \Nset^k$ such that
$\sum_{j=1}^k c_j a'_j \in [l',u']$.
%
\begin{table}[tb]
\begin{center}
\begin{tabular}[t]{rcr|M}
element & symbol & NN & \multicolumn{1}{c}{mass (Da)} \\

\hline

hydrogen & \ce{H} &  1 &  1.007825 \\
carbon   & \ce{C} & 12 & 12.000000 \\
nitrogen & \ce{N} & 14 & 14.003074 \\
oxygen   & \ce{O} & 16 & 15.994915 \\
phosphor & \ce{P} & 31 & 30.973762
\end{tabular}
\hspace{3ex}
\begin{tabular}[t]{rcr|M}
element & symbol & NN & \multicolumn{1}{c}{mass (Da)} \\

\hline

sulfur   & \ce{S}  &  32 &  31.972071 \\
chlorine & \ce{Cl} &  35 &  34.968853 \\
bromine  & \ce{Br} &  79 &  78.918337 \\
iodine   & \ce{I}  & 127 & 126.904473
\end{tabular}
\end{center}

\medskip
\caption{Elements considered in this paper.  For each element we report the
  \emph{monoisotopic mass}, that is, the mass of the naturally occurring
  isotope with smallest nucleon number (NN).  Masses taken
  from~\cite{audi03ame2003}.}
\label{tab:masses}
\end{table}
%
Analogously, we can decompose integer masses over an alphabet of integer
masses $a_1, \dots, a_k$.  Again, we assume that masses are ordered and
that $a_1$ is minimum.  We want to find all \emph{decompositions} $c = (c_1,
\dots, c_k) \in \Nset^k$ such that $\sum_{j=1}^k c_j a_j \in \{l,\dots,u\}$
where $l,u$ are integer.

B\"ocker \etal~\cite{boecker09sirius} describe how to transform an instance
of the real-valued mass decomposition problem into an integer-valued
instance, see there for details.  We briefly recapitulate the method: For a
given blowup factor $b \in \Rset$ we transform real-valued masses
$a'_1,\dots,a'_k$ into integer masses $a_j := \lfloor b a'_j \rfloor$.
(Different from~\cite{boecker09sirius} we will round down here, as this
presentation appears to be somewhat easier to follow.)  We want to find all
real-valued decompositions in the interval $[l',u'] \subseteq \Rset$.
Regarding the upper bound we have $\sum_j c_j a_j \le b \sum_j c_j a'_j \le b u'$
and, as the left side is integer, $\sum_j c_j a_j \le \lfloor b u' \rfloor$.
For the lower bound, we have to take into account rounding error
accumulation: We define relative rounding errors
\begin{equation} \label{equ:delta}
  \Delta_j = \Delta_j(b) := \frac{b a'_j - \lfloor b a'_j \rfloor}{a'_j} = b -
  \frac{\lfloor b a'_j \rfloor}{a'_j} \quad \text{for $j=1,\dots,k$,}
\end{equation}
and note that $0 \le \Delta_j < \frac{1}{a'_j}$.  Let $\Delta = \Delta(b) :=
\max\nolimits_j \{\Delta_j\}$.  Then, $\sum_j a'_j c_j \ge l'$
implies $\sum_j a_j c_j \ge b l' - \Delta l'$, see \cite{boecker09sirius} for
details.  To this end, we can decompose integer masses in the interval $l :=
\lceil b l' - \Delta l' \rceil$ to $u := \lfloor b u' \rfloor$.  Doing so, we
guarantee that no real-valued decomposition will be missed.  The list of
integer decompositions will contain false positive decompositions, but these
can be easily filtered out by checking $\sum_j c_j a'_j \in [l',u']$ for each
integer decomposition.

Mass accuracy of an MS instrument depends linearly on the mass that we
measure, and is usually given in ``parts per million'' (ppm).  Formally, for
a given mass $m \in \Rset$ and some $\epsilon > 0$, we want to find all
masses in the interval $[l',u']$ with $l' := (1-\epsilon) m$ and $u' :=
(1+\epsilon) m$.  To this end, the width $u'-l' = 2 \epsilon m$ of the
interval that we want to decompose, is \emph{linear} in the mass~$m$.

For integer masses, the number of decompositions $\gamma(m)$ of some mass $m$
asymptotically equals $\gamma(m) \sim \frac{1}{a_1 \cdots a_k}
m^{k-1}$~\cite{wilf94generatingfunctionology}.  This leads to a similar
estimate for real-valued masses~\cite{boecker09sirius}.  In general, this
asymptotic estimate is accurate only for very large masses; for molecular
formulas, it is a relatively good estimate even for small
masses~\cite{boecker09sirius}.


\section{Algorithms for decomposing masses}

The conceptually simplest algorithm for decomposing masses is a search tree
that recursively builds up the decompositions (molecular formulas), taking
into account the mass accuracy.  The algorithm is very similar to
\textsc{FindAllRecursive} in Fig.~\ref{fig:findall-rec}, we leave out the
straightforward details.  This algorithm has been suggested several times in
the literature~\cite{robertson77massform, dromey80calculation,
  agarwal12stoichiometry}.  The major disadvantage of this algorithm is that
its running time is not output-sensitive: For a constant alphabet of size
$k$, the algorithm requires $\Theta(m^{k-1})$ time, even if there is not a
single decomposition.
\begin{figure}[tb]
\begin{algorithmic}[1]
\Procedure{FindAllRecursive}{integer $i \le k$, mass $m$, decomposition $c$}
  \If {$i=0$} \State Output~$c$ and return
  \EndIf
  \If {$\decomposable(i-1,m) = 1$}
    \State $\textsc{FindAllRecursive} (i-1,m,c)$
  \EndIf
  \If {$m \ge a_i$ and $\decomposable(i,m-a_i) = 1$}
    \State $\textsc{FindAllRecursive} (i,m-a_i,c+e_i)$
  \EndIf
\EndProcedure
\end{algorithmic}

\caption{Recursive algorithm for enumerating all decompositions of a given
  mass~$m$.  To decompose mass $M$, this algorithm is initially called as
  $\textsc{FindAllRecursive} (k,M,0)$.  Vector $e_i$ denotes the
  $i$\textsuperscript{th} unit vector.}
\label{fig:findall-rec}
\end{figure}

To decompose an integer over an alphabet $\Sigma = \{a_1,\dots,a_k\}$ of
integer masses, we can use algorithm \textsc{FindAllRecursive} in
Fig.~\ref{fig:findall-rec}.  This algorithm requires an oracle such that
$\decomposable(i,m) = 1$ if and only if $m$ is decomposable over the
sub-alphabet $\{a_1,\dots,a_i\}$.  We can build this oracle using a dynamic
programming table $D[i,m] = \decomposable(i,m)$: We initialize $D[0,0] = 1$,
$D[0,m] = 0$ for $m \ge 1$, and use the recurrence $D[i,m] = \max \{
D[i-1,m], D[i,m-a_i] \}$ for $m \ge a_i$ and $D[i,m] = D[i-1,m]$ otherwise.
This approach requires $O(k M)$ memory to store $D$ and $O(k M)$ time to
compute it, where $M$ is the largest mass that we want to decompose.  The
algorithm has polynomial time and space with regards to the mass $m$ we want
to decompose.\footnote{Precisely speaking, running time is pseudo-polynomial
  in the input $m$, as polynomial running time would require polynomial
  dependency on~$\log m$.  Since the decision version of the problem (``is
  there a decomposition of mass~$m$?'') is weakly NP-hard~\cite{lueker75two},
  there is little hope for an algorithm with running time polynomial in~$\log
  m$.  We will ignore this detail in the following.}  Time for computing each
decomposition is $O(k m / a_1)$.  The decomposition algorithm has polynomial
delay and, hence, is output-dependent.

A more memory-efficient way to build the required oracle, is to use the
extended residue table (ERT) from~\cite{boecker05efficient}: For an integer $m$,
let $m \modulo a_1$ denote the \emph{residue} of $m$ modulo~$a_1$, where $m
\modulo a_1 \in \{0, \dots, a_1-1\}$ .  We define the \emph{extended residue
  table} $N[0 \dots k, 0 \dots a_1-1]$ by
\[
  N[i,r] = \min \bigl\{ m \WITH \text{$r = m \modulo a_1$, and $m$ is
    decomposable over $\{a_1,\ldots, a_i\}$} \bigr\}
\]
where we define $N[i,r] = +\infty$ if no such number exists, that is, if the
minimum is taken over the empty set.  Now, we can define the oracle by
\begin{equation}\label{eq:oracle}
  \decomposable(i,m) := \begin{cases}
    1 & \text{if $m \ge N[i, m \modulo a_1]$,} \\
    0 & \text{otherwise.}
  \end{cases}
\end{equation}
Storing the ERT requires $O(k a_1)$ space, and the table can be
computed in $O(k a_1)$ time.  The time for computing each decomposition using
algorithm \textsc{FindAllRecursive} is again $O(k m / a_1)$ but can be
reduced to $O(k a_1)$~\cite{boecker05efficient, boecker07fast}.  The
conceptual advantage of this approach is that we do not have to decide upon
some ``largest mass'' during preprocessing, and that both time and space do
no longer depend on the mass $m$ that we want to decompose.


\section{Iterative version of the decomposition algorithm}

The \naive search tree algorithm for decomposing masses can easily be made
iterative using $k$ nested \textsc{For}-loops.  Replacing algorithm
\textsc{FindAllRecursive} by an iterative version is slightly more
complicated, as we have to avoid ``empty branches'' of the search tree, where
no decomposition can be found.  We can use an auxiliary Boolean vector $d[1
  \dots k]$ that stores which of the two alternative recursive calls from
\textsc{FindAllRecursive} has been executed last.  We present a more
involved version of the iterative algorithm in Fig.~\ref{fig:findall-it}.  We
avoid the auxiliary vector by deciding on the alternative calls directly from
the decomposition~$c$.

\begin{figure}[tb]
\begin{algorithmic}[1]
  \Procedure{FindAllIterative}{mass $m$}
  \State decomposition $c = (c_1,\dots,c_k) \gets 0$
  \State integer $i \gets k$
  \Comment constant alphabet size $k$
  \While {$i \le k$}
    \If{$\decomposable(i,m)=0$} \label{line:entry-while}
      \Comment is this decomposable at all?
      \While {$i \le k$ and $\decomposable(i,m) = 0$}
        \Comment no, end ``recursion''
        \State $m \gets m + c_i a_i$
        \State $c_i \gets 0$
        \State $i \gets i+1$
      \EndWhile
      \Comment now, $\decomposable(i,m)=1$ holds
      \If {$i \le k$}
        \State $m \gets m-a_i$
        \State $c_i \gets c_i+1$
      \EndIf
    \Else
    \Comment yes, decomposable
      \While {$i > 1$ and $\decomposable(i-1,m)=1$}
        \State $i \gets i-1$
      \EndWhile
      \Comment now, $\decomposable(i,m)=1$
      \If {$i=1$}
      \Comment output decomposition
        \State $c_1 \gets \lfloor m / a_1 \rfloor$
        \Comment $(\ast)$
        \State Output~$c = (c_1,\dots,c_k)$
        \State $i \gets 2$
        \Comment correct $i$
      \EndIf

      \If {$i \le k$}
      \Comment move to next element
        \State $m \gets m-a_i$
        \State $c_i \gets c_i+1$
      \EndIf
    \EndIf
  \EndWhile
\EndProcedure
\end{algorithmic}

\caption{Iterative algorithm for enumerating all decompositions of a given
  mass~$m$.}
\label{fig:findall-it}
\end{figure}

The algorithm is independent of the actual implementation of the oracle
$\decomposable(i,m)$.  Asymptotically, worst-case running time is identical to
that of the recursive version; in practice, the iterative version is
nevertheless considerably faster than its recursive counterpart, as we avoid
the stack handling.


\section{Selecting optimal blowup factors}

Transforming the real-valued decomposition problem into its integer-valued
counterpart requires that we choose some blowup factor $b \in \Rset$.  Due to
the rounding error correction, we have to decompose roughly $\Delta(b) u$
``auxiliary'' integers in addition to the $b (u-l+1)$ ``regular'' integers,
where $(u-l+1) \in \Theta(u)$.  It is reasonable to ask for a blowup factor
such that the ratio of additional integers $\frac{\Delta(b) u}{b u} =
\frac{\Delta(b)}{b}$ is minimum.  For ``sufficiently small'' $b > 0$ we have
$\Delta(b) = b$ and, hence, $\frac{\Delta(b)}{b} = 1$.

Since $\Delta(b) < \max\nolimits_j \{ \frac{1}{a'_j} \}$ is bounded, we can
make $\frac{\Delta(b)}{b}$ arbitrarily small by choosing an arbitrarily large
blowup factor~$b$.  But this is not realistic in applications, as memory
requirements increase linearly with~$b$.  To this end, we suppose that memory
considerations imply an upper bound of $B \in \Rset$.  We want to find $b \in
(0,B)$ such that $\frac{\Delta(b)}{b}$ is minimized.  We can explicitly find
an optimal $b$ as follows: First, we consider the functions
\[
  \Delta_j: \Rset \to \Rset \quad \text{with} \quad \Delta_j(b) := b -
  \tfrac{1}{a'_j} \lfloor b a'_j \rfloor,
\]
for all $j=1,\dots,k$.  Each $\Delta_j$ is a piecewise linear function with
discontinuities $\frac{1}{a'_j}, \frac{2}{a'_j}, \dots, \frac{\lfloor a'_j B
  \rfloor}{a'_j}$.  In every interval, this function has slope~$1$.  Next, we
set $\phi_1 \equiv \Delta_1$ and for $j \ge 2$, we define $\phi_j$ as the
maximum of $\phi_{j-1}$ and $\Delta_j$.  Then, $\phi_j$ is a piecewise linear
function with $O \bigl( (a'_1 + \dots + a'_j) B \bigr)$ discontinuities.
Finally, $\Delta \equiv \phi_k$ is a piecewise linear function with $O \bigl(
(a'_1 + \dots + a'_k) B \bigr)$ discontinuities.  We sweep over the discontinuities
from left to right, and for each discontinuity $b$, we calculate all
$\Delta_j(b)$ and $\Delta(b)$.  This can be easily achieved in time $O(k
(a'_1 + \dots + a'_k) B) = O(k^2 a'_k B)$, where $a'_k$ is the largest mass
in the alphabet.  For every piecewise linear part $I \subseteq \Rset$ of
$\Delta$ the minimum of $\frac{\Delta(b)}{b}$ must be located at one of the
terminal points, so it suffices to test the $O(k a'_k B)$ discontinuities to
find the minimum of $\frac{\Delta(b)}{b}$.

\begin{table}[h]

\centering
\begin{scriptsize}

\begin{tabular}[t]{d{8}d{7}d{12}}
\multicolumn{1}{c}{blowup $b$} & \multicolumn{1}{c}{$\Delta(b)$} &
\multicolumn{1}{c}{$\Delta(b)/b$} \\

\hline

 1127.1810743 & 0.014407 & 1.278199 \cdot 10^{-5} \\
 1128.1808548 & 0.014188 & 1.257608 \cdot 10^{-5} \\
 1181.7527469 & 0.012223 & 1.034371 \cdot 10^{-5} \\
 1182.7510330 & 0.010729^{\ast} & 9.071515 \cdot 10^{-6} \\
 1680.8473159 & 0.013982 & 8.318731 \cdot 10^{-6} \\
 1681.8470521 & 0.013718 & 8.156909 \cdot 10^{-6} \\
 1896.1666667 & 0.013377 & 7.054863 \cdot 10^{-6} \\
 1897.1666667 & 0.012530 & 6.604628 \cdot 10^{-6} \\
 2064.8444567 & 0.012121 & 5.870279 \cdot 10^{-6} \\
 2309.9247647 & 0.008097 & 3.505730 \cdot 10^{-6} \\
 2310.9237056 & 0.007038^{\ast}  & 3.045939 \cdot 10^{-6} \\
 2939.0036991 & 0.007079 & 2.408730 \cdot 10^{-6} \\
 5248.9269781 & 0.010311 & 1.964477 \cdot 10^{-6} \\
 5334.2592503 & 0.009867 & 1.849794 \cdot 10^{-6} \\
 5335.2582387 & 0.008238 & 1.544184 \cdot 10^{-6} \\
 5963.3376861 & 0.008003^{\ast} & 1.342117 \cdot 10^{-6} \\
 8519.3436784 & 0.010925 & 1.282415 \cdot 10^{-6} \\
 9072.0116320 & 0.011631 & 1.282179 \cdot 10^{-6} \\
 9456.0064464 & 0.011287 & 1.193667 \cdot 10^{-6} \\
 9457.0057796 & 0.010840 & 1.146246 \cdot 10^{-6} \\
 9701.0919315 & 0.009977 & 1.028442 \cdot 10^{-6} \\
10415.5000000 & 0.006917^{\ast} & 6.641097 \cdot 10^{-7} \\
12725.4231558 & 0.007214 & 5.669173 \cdot 10^{-7} \\
12726.4199232 & 0.004531^{\ast} & 3.560891 \cdot 10^{-7} \\
18689.7544746 & 0.004911 & 2.627644 \cdot 10^{-7} \\
26080.9191881 & 0.005617 & 2.153888 \cdot 10^{-7} \\
29105.2521655 & 0.005901 & 2.027566 \cdot 10^{-7} \\
32044.2526951 & 0.005370^{\ast} & 1.676031 \cdot 10^{-7} \\
42459.7510861 & 0.006746 & 1.588932 \cdot 10^{-7} \\
42460.7500000 & 0.006678 & 1.572787 \cdot 10^{-7} \\
44770.6696249 & 0.002958^{\ast} & 6.607444 \cdot 10^{-8} \\
96687.4182692 & 0.005238 & 5.417847 \cdot 10^{-8} \\
\end{tabular}
\hspace{3ex}
\begin{tabular}[t]{d{8}d{7}d{12}}
\multicolumn{1}{c}{blowup $b$} & \multicolumn{1}{c}{$\Delta(b)$} &
\multicolumn{1}{c}{$\Delta(b)/b$} \\

\hline

 1127.1810743 & 0.014407 & 1.278199 \cdot 10^{-5} \\
 1128.1808548 & 0.014188 & 1.257608 \cdot 10^{-5} \\
 1182.7510330 & 0.012772^{\ast} & 1.079876 \cdot 10^{-5} \\
 1680.8473159 & 0.013982 & 8.318731 \cdot 10^{-6} \\
 1681.8470521 & 0.013718 & 8.156909 \cdot 10^{-6} \\
 2064.8444567 & 0.012121 & 5.870279 \cdot 10^{-6} \\
 2309.9247647 & 0.011976 & 5.184759 \cdot 10^{-6} \\
 2310.9237056 & 0.010026^{\ast} & 4.338805 \cdot 10^{-6} \\
 3268.4252033 & 0.013661 & 4.179833 \cdot 10^{-6} \\
 3269.4249838 & 0.012594^{\ast} & 3.852308 \cdot 10^{-6} \\
 3897.5019225 & 0.013159 & 3.376302 \cdot 10^{-6} \\
 3898.5011421 & 0.012060 & 3.093702 \cdot 10^{-6} \\
 4206.0872326 & 0.011084 & 2.635363 \cdot 10^{-6} \\
 4207.0871650 & 0.010920^{\ast} & 2.595644 \cdot 10^{-6} \\
 5248.9282869 & 0.011620 & 2.213814 \cdot 10^{-6} \\
 5802.5956469 & 0.012587 & 2.169277 \cdot 10^{-6} \\
 5963.3376861 & 0.008789^{\ast} & 1.473957 \cdot 10^{-6} \\
 7146.0837847 & 0.010040 & 1.405093 \cdot 10^{-6} \\
 9701.0919315 & 0.009977 & 1.028442 \cdot 10^{-6} \\
10415.5007612 & 0.007678^{\ast} & 7.371910 \cdot 10^{-7} \\
15664.4258530 & 0.009633 & 6.149738 \cdot 10^{-7} \\
16378.8350902 & 0.008707 & 5.316052 \cdot 10^{-7} \\
16379.8339400 & 0.006683^{\ast} & 4.080188 \cdot 10^{-7} \\
26710.0000000 & 0.010596 & 3.967193 \cdot 10^{-7} \\
26794.3334813 & 0.010397 & 3.880516 \cdot 10^{-7} \\
26795.3333334 & 0.009376 & 3.499167 \cdot 10^{-7} \\
28390.8415041 & 0.008170 & 2.877948 \cdot 10^{-7} \\
29105.2521655 & 0.007871^{\ast} & 2.704370 \cdot 10^{-7} \\
34355.1749622 & 0.008295 & 2.414630 \cdot 10^{-7} \\
38807.3390692 & 0.008710 & 2.244429 \cdot 10^{-7} \\
44769.6758508 & 0.009184 & 2.051405 \cdot 10^{-7} \\
44770.6721964 & 0.005529^{\ast} & 1.235112 \cdot 10^{-7} \\
63460.4230255 & 0.006358 & 1.002015 \cdot 10^{-7} \\
90170.4178028 & 0.008375 & 9.288531 \cdot 10^{-8} \\
96687.4182692 & 0.006542 & 6.767102 \cdot 10^{-8} \\
\end{tabular}
\end{scriptsize}

\medskip
\caption{Locally optimal blowup factors in the range $b \in [1000, 100000]$,
  for alphabet of elements \ce{CHNOPS} (left) and \ce{CHNOPSClBrI} (right).
  For two consecutive entries $b',b''$ from the table, $b'$ is locally
  optimal in $(0,b'')$, so $\frac{\Delta(b')}{b'} \le \frac{\Delta(b)}{b}$
  for all $b \in (0,b'')$.  Values $b$ rounded up, other values rounded down.
  $^{\ast}$Entries $\Delta(b)$ that are smaller than both the previous and
  the following entry.}
\label{tab:blowup}
\end{table}

We compute optimal blowup factors for the default alphabet \ce{CHNOPS}, and
for the extended alphabet \ce{CHNOPSClBrI} suggested
in~\cite{stravs13automatic}.  Since we can find arbitrarily small blowup
factors by increasing $b$, any blowup factor $b' \in \Rset$ can only be
\emph{locally optimal}: that is, for an upper bound $b'' \in \Rset$ and all
$b \in (0,b'')$ we then have $\frac{\Delta(b')}{b'} \le \frac{\Delta(b)}{b}$.
See Table~\ref{tab:blowup} for all locally optimal blowup factors in the
range $b \in [1000, 100000]$.  We do not report blowup factors below $1000$
as, for the mass accuracies considered here, such blowup factors result in a
dramatic increase of false positive decompositions and, hence, are not useful
in practice.


\section{Range decompositions} \label{sec:range}

Agarwal \etal~\cite{agarwal12stoichiometry} suggested to decompose a range of
masses $m, \dots, m+\mu-1$ for integers $m,\mu$, instead of decomposing each
mass individually.  In theory, this does not noticeably improve running
times: Using the approaches described above, we can iterate over all masses
$m' = m, \dots, m+\mu-1$.  Let $\gamma(m,m+\mu)$ denote the number of
decompositions in this range, then this results in a total running time of
$O(\gamma(m,m+\mu) k a_1 + \mu)$ for the approach
of~\cite{boecker05efficient, boecker07fast}.  Clearly, the additive $O(\mu)$
term can be ignored in practice.

But from an algorithm engineering perspective, decomposing a range instead of
an integer may result in considerable time savings: For the algorithm
\textsc{FindAllRecursive}, this can significantly reduce the number of
recursive function calls.  To this end, given a range $m, \dots, m+\mu-1$ and
an alphabet $\Sigma = \{a_1,\ldots,a_k\}$ we assume an oracle with
$\decomposable(i,m) = 1$ if and only if there is at least one $m' \in \{m,
\dots, m+\mu-1\}$ that is decomposable over $\{a_1,\ldots,a_i\}$.  Solely for
the sake of clarity, we will assume $\mu$ to be fixed, although it obviously
depends on the mass that we want to decompose.  With this new oracle, we can
reuse the algorithms from Fig.~\ref{fig:findall-rec} and~\ref{fig:findall-it}
without further changes.

In the following, let $\decomposable_0(i,m)$ denote the original oracle for a
single mass~$m$.  Then, a straightforward oracle for the range decomposition
is
\[
  \decomposable(i,m) = \max_{m' \in \{m\ldots m+\mu-1\}}
  \decomposable_0(i,m') .
\]
But this requires $\mu$ calls of the $\decomposable_0(i,m)$ oracle and
results in a \emph{multiplicative} factor of $O(\mu)$ in the running time.
Agarwal \etal~\cite{agarwal12stoichiometry} suggested modifying the integer
knapsack recurrence mentioned above, to capture the mass range: To this end,
we initialize $D_{\mu}[0,m] = 1$ for $m=0,\dots,\mu-1$ and $D_{\mu}[0,m] = 0$
for $m \ge \mu$.  We use the same recurrence as above, namely $D_{\mu}[i,m] =
\max \{ D_{\mu}[i-1,m], D_{\mu}[i,m-a_i] \}$ for $m \ge a_i$ and
$D_{\mu}[i,m] = D_{\mu}[i-1,m]$ otherwise.  Unfortunately, for each $\mu$
that we want to consider for decomposing, this requires preprocessing and
storing a dynamic programming table.  Again, this is not desirable in
application, as the mass error of the measurement increases with mass.  So,
we have to compute and store a table for every $\mu = 1,\dots,\mumax$.

But with a small trick, we can reduce the multiplicative factor for space
from $O(\mumax)$ to $O(\log \mumax)$: Let $\decomposable_l$ be an oracle with
$\decomposable(i,m) = 1$ if and only if there is at least one $m' \in \{m,
\dots, m + 2^l -1\}$ that is decomposable over $\{a_1,\ldots,a_i\}$.  Then,
we can ``recover'' the oracle $\decomposable$ for the range $\{m, \dots, m
+ \mu -1\}$ as
\[
  \decomposable(i,m) = \max \bigl\{ \decomposable_l (i,m), \decomposable_l
  (i, m+\mu-2^l) \bigr\}
\]
where $l := \lfloor \log_2 \mu \rfloor$.

We will now show how to use the extended residue table
from~\cite{boecker05efficient} for range decompositions: We define a family
of extended residue tables $N_l$ for $l = 0,\dots,\lfloor \log_2 \mu
\rfloor$, where $N_l[i,r]$ is the minimum of all $m$ with $r = m \modulo
a_1$, such that some $m' \in \{m,\dots,m+2^l-1\}$ is decomposable over
$\{a_1,\ldots, a_i\}$.  Again, $N_l[i,r] = +\infty$ if no such number exists.
Now, we can re-use the oracle from~\eqref{eq:oracle}: We have
$\decomposable_l(i,m) = 1$ if and only if $m \ge N_l[i,m]$.  Storing all
extended residue tables requires $O(k a_1 \log \mumax)$ space, and the tables
can be computed in $O(k a_1 \log \mumax)$ time using the following simple
recurrence: We initialize $N_0[i,r] = N[i,r]$ and use
\begin{equation}
  N_{l+1}[i,r] = min \bigl\{ N_l[i,r], N_l[i, (r + 2^l) \modulo a_1] \bigr\}
\end{equation}
for $l = 0,\dots,\mumax-1$, $i = 0,\dots,k$, and $r = 0, \dots, a_1-1$.
Here, $N[i,r]$ refers to the extended residue table for the single integer
decomposition problem.

The iterative algorithm \textsc{FindAllIterative} does not consider the
degenerate case where the width of the interval we want to decompose, is
large compared to the masses of the alphabet.  In particular, for $u-l \ge
a_1$ every decomposition with mass at most $u$ can be ``completed'' using
element $a_1$ to find a decomposition with mass in $\{l,\dots,u\}$.  For this
case, we have to adapt the algorithm by replacing the line marked $(\ast)$ in
Fig.~\ref{fig:findall-it} by a loop over appropriate numbers of
elements~$a_1$.  But for this degenerate case, we find a decomposition for
every leaf of the \naive search tree algorithm; so, the iterative version of
this algorithm outperforms all other, more involved algorithms.


\section{Results}

We implement the algorithms mentioned above in Java~1.6.
\textsc{SearchTree} denotes the \naive search tree algorithm.  We distinguish
two algorithms based on the recursive decomposition
(Fig.~\ref{fig:findall-rec}), namely \textsc{Recursive+Knapsack} using the
knapsack DP, and \textsc{Recursive+ERT} using the extended residue table.  We
also implement two versions of the iterative decomposition
(Fig.~\ref{fig:findall-it}), namely \textsc{Iterative+ERT} and
\textsc{Iterative+Range} which uses range decompositions from
Sec.~\ref{sec:range}.  For brevity, we exclude other combinations, as these
will in all likelihood not result in better running times.

\begin{table}[b]
\centering
\begin{tabular}{lC|RRRR}
& & \text{Orbitrap} & \text{MassBank} & \text{Eawag} & \text{Hill} \\ \hline

peaks & & 5\,393 & 2\,455 & 10\,017 & 12\,054 \\

maximum mass & & 1153 & 821 & 444 & 610 \\

median mass & & 205 & 211 & 149 & 186 \\
\end{tabular}

\medskip
\caption{Statistics of the datasets used in our evaluation.}
\label{tab:datasets}
\end{table}

We evaluate the algorithms on four datasets: The \emph{Orbitrap}
dataset~\cite{rasche12identifying} contains 97 compounds measured on a Thermo
Scientific Orbitrap XL instrument.  The \emph{MassBank}
dataset~\cite{horai10massbank} consists of 370 compounds measured on a Waters
Q-Tof Premier spectrometer.  The \emph{Eawag}
dataset~\cite{stravs13automatic} contains 60 compounds measured on a LTQ
Orbitrap XL Thermo Scientific and is also accessible from MassBank.  The
\emph{Hill} dataset~\cite{hill08mass} consists of 102 compounds with 502
spectra measured on a Waters Micromass QTOF II instrument.  We omit
experimental details.  All of these datasets are used in computations that
require the decomposition of peak masses.  See Table~\ref{tab:datasets} for
details.

We discard peaks with mass below $100$~Da because for such masses, the
problem becomes easy regardless of the used algorithm.  We report running
times for decomposing a single peak mass as well as all peaks in a dataset.
For algorithms working on integer masses, we generate integer-valued
instances as described in Sec.~\ref{sec:prelim}.  We use a mass accuracy of
20~ppm, so $\epsilon = 0.00002$.  For \textsc{Recursive+Knapsack},
\textsc{Recursive+ERT}, and \textsc{Iterative+ERT}, we decompose intervals
by decomposing all integer values separately.

\begin{table}[b]
\centering
\begin{tabular}{lC|RRRR}
algorithm & \text{blowup} & \text{Orbitrap} & \text{MassBank} & \text{Eawag}
& \text{Hill} \\ \hline

\textsc{SearchTree} & - & 166.0 \, \text{min} & 37.3 \, \text{min} & 2.3 \,
\text{min} & 23.8 \, \text{min} \\[1ex]

\textsc{Recursive+knapsack} & \multirow{2}{*}{100000} & 729.70 & 86.48 &
8.54 & 34.12  \\

\textsc{Recursive+ERT} & & 308.22 & 39.67 & 2.10 & 19.85  \\[1ex]

\multirow{5}{*}{\textsc{Iterative+ERT}} & 100000 & 216.21 & 30.83 & 2.10 &
14.64 \\

 & 44770.6721964 & 129.97 & 21.25 & 1.62 & 9.75   \\ 

 & 5963.3376861 & 102.97 & 17.92 & 0.94 & 7.32   \\ 

 & 1182.7510330 & 122.91 & 22.40 & 0.95 & 7.66   \\ 

 & 1000 & 2289.36 & 420.58 & 15.81 & 190.32   \\[1ex]

\textsc{Iterative+range} & 5963.3376861 & 66.88 & 13.98 & 0.88 & 6.13 \\

\\

algorithm & \text{blowup} & \text{Orbitrap} & \text{MassBank} & \text{Eawag}
& \text{Hill} \\ \hline

\textsc{SearchTree} & - & 221.97 & 106.16 & 28.65 & 131.07
\\[1ex]

\textsc{Recursive+knapsack} & \multirow{2}{*}{100000} & 54.47 & 19.98 &
5.62 & 13.21  \\

\textsc{Recursive+ERT} & & 18.03 & 5.77 & 1.17 & 4.77 \\[1ex]

\multirow{5}{*}{\textsc{Iterative+ERT}} & 100000 & 12.99 & 4.63 & 1.15 &
4.31 \\

 & 44770.6721964 & 10.15 & 3.63 & 0.75 & 3.14   \\ 

 & 5963.3376861 & 5.86  & 2.27 & 0.39 & 1.93   \\ 

 & 1182.7510330 & 7.39 & 2.76 & 0.40 & 2.17   \\ 

 & 1000 & 121.83 & 46.60 & 5.89 & 40.71  \\[1ex]

\textsc{Iterative+range} & 5963.3376861 & 4.71 & 1.89 & 0.36 & 1.77
\end{tabular}

\medskip
\caption{We report running times of the algorithms for decomposing all peaks
  in a dataset, for alphabets \ce{CHNOPSClBrI} (top) and \ce{CHNOPS}
  (bottom).  Running times are reported in seconds except for \textsc{SearchTree}
  and alphabet \ce{CHNOPSClBrI}, where running times are reported in
  minutes. For all measurements, we use mass accuracy $20$~ppm.}
\label{tab:results}
\end{table}

\begin{figure}[tb]
\centering
\includegraphics[width=0.48\textwidth, trim = 0 15 15 50, clip]%
{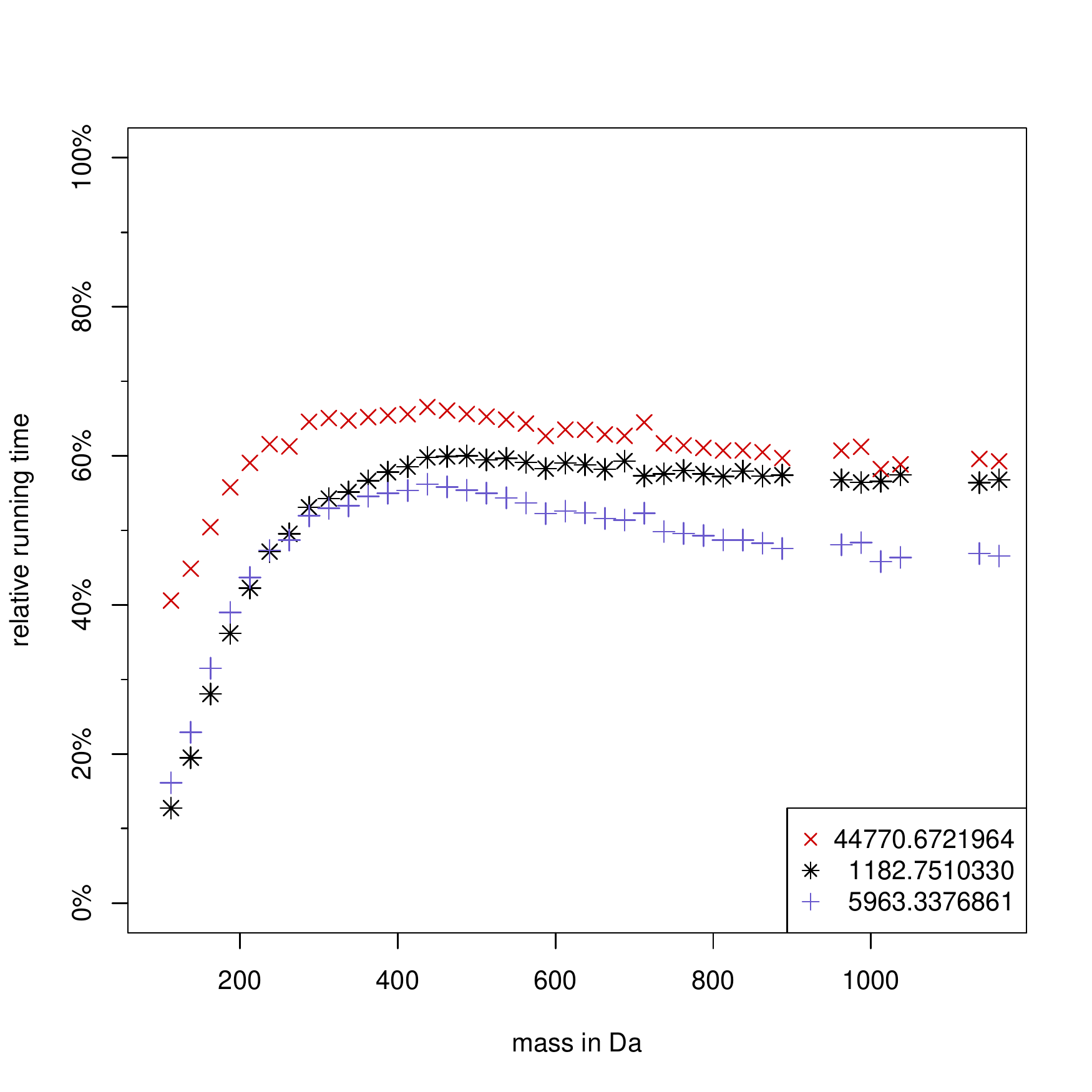}
\hspace{0.01\textwidth}
\includegraphics[width=0.48\textwidth, trim = 0 15 15 50, clip]%
{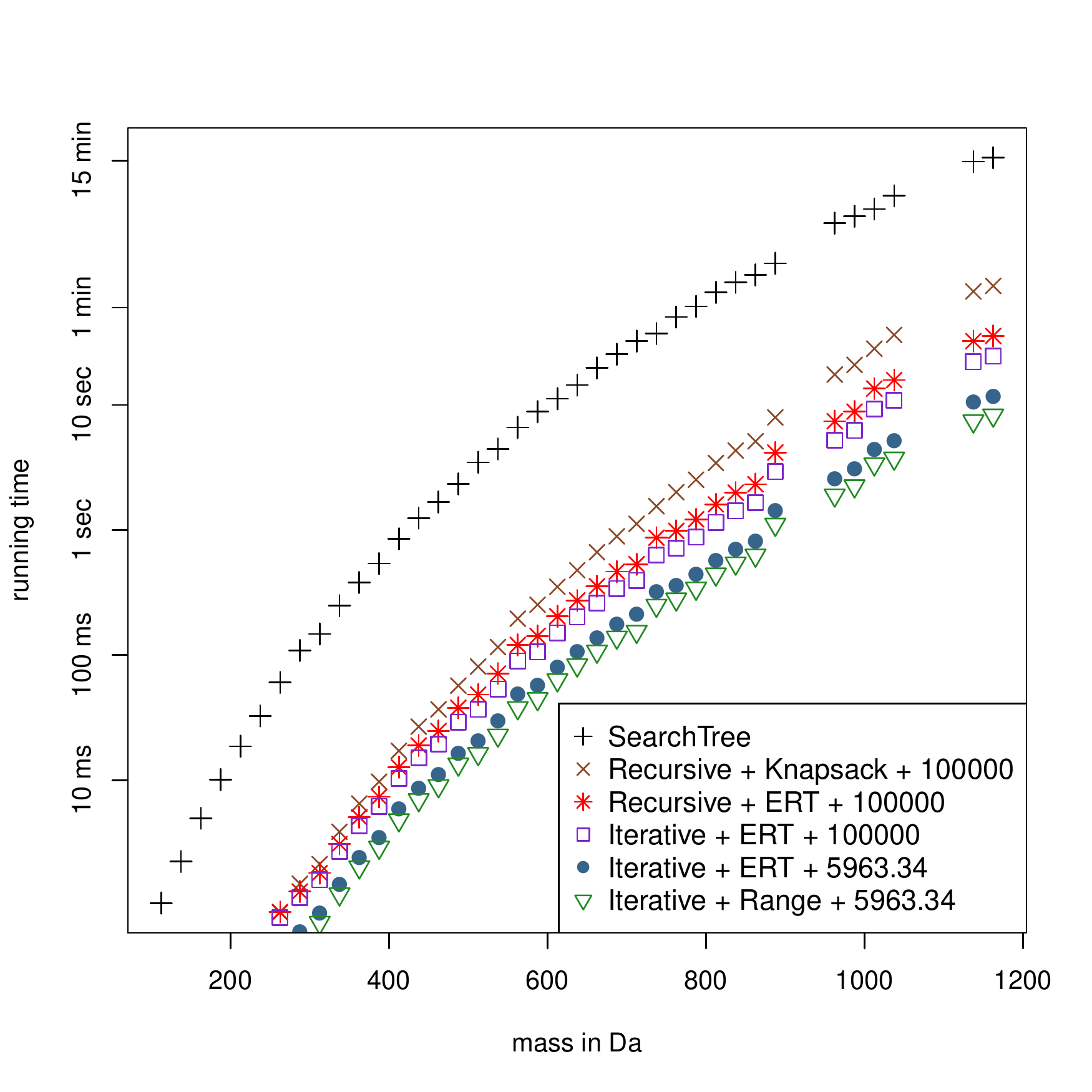}

\includegraphics[width=0.48\textwidth, trim = 0 15 15 50, clip]%
{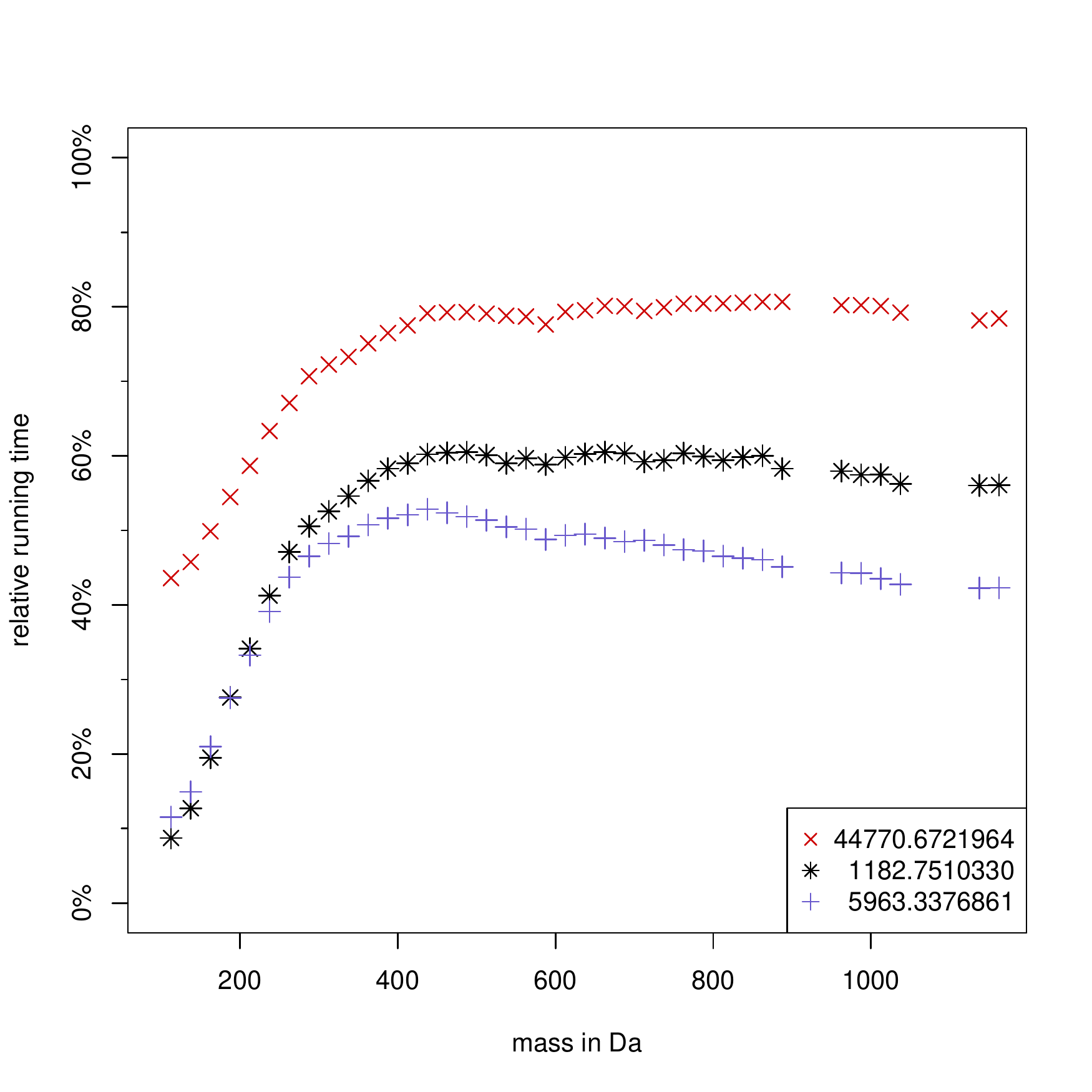}
\hspace{0.01\textwidth}
\includegraphics[width=0.48\textwidth, trim = 0 15 15 50, clip]%
{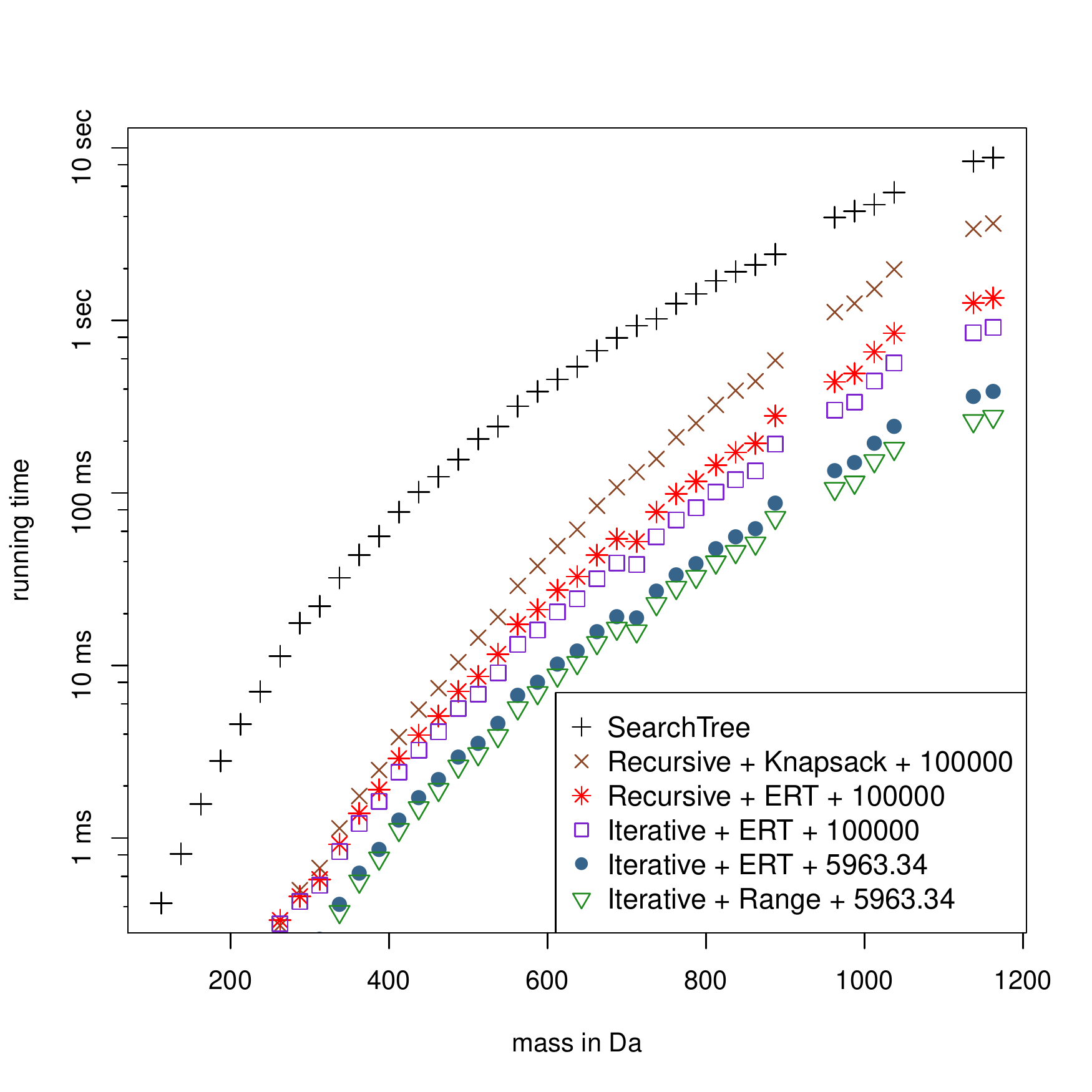}

\caption{Running times on the Orbitrap dataset using alphabet
  \ce{CHNOPSClBrI} (top) or \ce{CHNOPS} (bottom), and $20$~ppm mass accuracy.
  Average running times are reported for bins of width $25$~Da.  Left: Relative
  running times of \textsc{Iterative+ERT} for different blowup factors,
  normalized to blowup factor $b=100000$ as $100\,\%$.  Right: Running times
  for different algorithms.  Note the logarithmic y-axis.}
\label{fig:CHNOPSExt}
\end{figure}

Running time measurements are done on a Intel Xeon E5645 with 48 GB RAM.
For each algorithm we repeat computations five times and report minimum
running times.  For the complete datasets, total running times can be found
in Table~\ref{tab:results} and Figure~\ref{fig:CHNOPSExt} for alphabets
\ce{CHNOPSClBrI} and \ce{CHNOPS}.  We find that the fastest ERT-based
algorithm \textsc{Iterative+Range} was $56$-fold faster than the
\textsc{SearchTree} algorithm for alphabet \ce{CHNOPS}; this increases to
$150$-fold speedup for alphabet \ce{CHNOPSClBrI}.

Replacing the knapsack DP by an ERT table results in a $2.3$-fold speedup.
In addition, memory requirements decrease considerably: For example, to
decompose the maximal mass of $1153.395$ with a blowup of $100\,000$ and the
extended alphabet, the integer knapsack DP table requires $124$ megabyte,
whereas the ERT requires only $3.5$ megabyte.

Replacing the recursive search by its iterative counterpart has only limited
impact: We find that the iterative algorithm \textsc{Iterative+ERT} is
merely $1.4$-fold faster than its recursive counterpart,
\textsc{Recursive+ERT}.

We observe that the blowup factor has a major impact on running times.
Choosing
locally optimal blowup factors does in fact significantly reduce running
times: For example, blowup factor $b=1182.7510330$ results in $19$-fold faster
running times than $b=1000$.  We test all locally optimal blowup factors
from Table~\ref{tab:blowup}. 
We observe best running time for blowup $b = 
5963.3376861$, whereas larger and smaller blowup factors result in increased
running times.  We refrain from reporting all running times, see
Fig.~\ref{fig:CHNOPSExt} and Table~\ref{tab:results} for some examples.  The
best blowup factor $b = 5963.3376861$ results in a two-fold speedup when
compared to the ``default'' blowup factor $b=100000$
from~\cite{boecker09sirius}.  As a pleasant side effect, this decreases the
memory requirements of the algorithm by $94\,\%$.

The range decomposition improves running times by about $1.5$-fold for the
best blowup factor.  A stronger improvement is achieved when more integers
are to be decomposed using, say, a larger blowup factor.  Using blowup
factor $b = 5963.3376861$, memory increases to $2.1$~MB for
storing ten ERT tables.

Finally, we repeat our experiments for an improved mass accuracy $1$~ppm.
For all algorithms except \textsc{SearchTree} this results in roughly a
$6$- to $11$-fold decrease of running time, whereas \textsc{SearchTree} running
times does not change.  For this mass accuracy and alphabet
\ce{CHNOPSClBrI}, the best algorithm \textsc{Iterative+range} is
$1000$-fold faster than the \naive \textsc{SearchTree} algorithm.


\section{Conclusion}

We suggest three techniques to improve the running time for decomposing real
masses.  We measure the improvements on four different datasets. All
techniques together result in a $4$-fold improvement in running time,
compared to the \textsc{Recursive+ERT} algorithm
from~\cite{boecker05efficient, boecker07fast}.  We note in passing that the
implementation of the \textsc{Recursive+ERT} algorithm used in this
evaluation, was two-fold faster than the one provided as part of
SIRIUS~\cite{boecker09sirius}.  The competitive edge of the new method is
even larger for ``hard'' problem instances, e.g. high masses, large mass
deviations, and bigger alphabets.  Compared to the \naive search tree
algorithm, we reach improvements between 56-fold and 1000-fold,
reducing the total running times from hours to minutes or even seconds.

Regarding the degenerate case $u-l \ge a_1$ mentioned in
Sec.~\ref{sec:range}, we argue that this case is of no interest, from either
the practical or the theoretical side: Modern MS instruments easily
reach mass accuracies of $10$~ppm and below, whereas metabolite and even
peptide masses rarely exceed $5000$~Da.  Even a peptide mass of $5000$~Da can
be measured with an accuracy of at least $0.05$~Da, well below the mass of a
single \ce{^1H} atom.  From the theoretical side, we would have to deal with a
humongous number of decompositions, rendering time to compute the
decompositions irrelevant in comparison to subsequent analysis steps.

\paragraph{Acknowledgments.}

We thank Tim White for proofreading earlier versions of this work.


\bibliographystyle{abbrv}
\bibliography{faster_decomp_wabi}

\end{document}